\documentclass[]{spie}  
\usepackage[]{graphicx}

\usepackage{amsfonts}
\usepackage{amsmath}
\usepackage{amssymb}
\usepackage{color}
\usepackage{cancel}

\newcommand{\bS}{\mathbf{S}}
\newcommand{\bL}{\mathbf{L}}

\newcommand{\T}{^{\text{T}}}

\newcommand{\beq}{\begin{equation}}
\newcommand{\eeq}{\end{equation}}

\title{Optical modular arithmetic} 

\author{Dmitri S. Pavlichin\supit{a} and Hideo Mabuchi\supit{a}
\skiplinehalf
\supit{a}Edward L.\ Ginzton Laboratory, Stanford University, Stanford, CA 94305
}

\authorinfo{Further author information: (Send correspondence to D.S.P.)\\D.S.P.: E-mail: dmitrip@stanford.edu\\  H.M.: E-mail: hmabuchi@stanford.edu}

\begin{document} 
  \maketitle 

\begin{abstract}
Nanoscale integrated photonic devices and circuits offer a path to ultra-low power computation at the few-photon level.  Here we propose an optical circuit that performs a ubiquitous operation: the controlled, random-access readout of a collection of stored memory phases or, equivalently, the computation of the inner product of a vector of phases with a binary ``selector'' vector, where the arithmetic is done modulo 2pi and the result is encoded in the phase of a coherent field.  This circuit, a collection of cascaded interferometers driven by a coherent input field, demonstrates the use of coherence as a computational resource, and of the use of recently-developed mathematical tools for modeling optical circuits with many coupled parts.  The construction extends in a straightforward way to the computation of matrix-vector and matrix-matrix products, and, with the inclusion of an optical feedback loop, to the computation of a ``weighted'' readout of stored memory phases.  We note some applications of these circuits for error correction and for computing tasks requiring fast vector inner products, e.g. statistical classification and some machine learning algorithms.
\end{abstract}







\section{Introduction}

Integrated nanoscale photonic circuits potentially offer significant improvements over eletronic circuits in terms of power consumption, interconnect density, and heat dissipation \cite{Beau11,Mill09}.   Circuits of photonic devices, involving the interaction of components coupled and powered by coherent fields, additionally present new computational resources \footnote{In the form of distributed phase coherence, even in the absence of entanglement} and tradeoffs.  Recently developed mathematical and software tools for modeling interconnected quantum optical systems have enabled the rapid exploration of the computational potential of such devices.  In a previous study \cite{PavlichinMabuchi13} we described a circuit architecture for decoding a class of error-correcting codes.

In this work we extend a growing toolbox of optical circuit motifs with a circuit that computes the inner product of two vectors --- one a binary selector vector and one an arbitrary vector of phases --- and encodes the output in the phase of a coherent field; the arithmetic is thus done modulo $2 \pi$.  We further describe a straightforward extension to the computation of matrix-vector and matrix-matrix products.  This operation could be used in the implementation of an optical random access memory (using a binary selector vector to read out the sum of an arbitrary subset of memory bits) or for one of many computing tasks involving binary vector inner products with modular arithmetic (for, e.g., linear error-correcting codes over a binary channel).  We further propose an alternate construction that involves an optical feedback loop, but yields a simpler model for the trouble.  The feedback construction can also use a non-binary ``selector'' weight, thus broadening the set of possible applications of our circuit (to, e.g., the computation of vector inner products as used in machine learning algorithms).

The circuit consists of a collection of Mach-Zehnder interferometers cascaded in series.  The ``memory bits'' are encoded in phases imparted upon an incident coherent field, while the binary selector bits are encoded in the binary settings of ``control'' phases ($0$ or $\pi$).  For the circuit to be useful, we assume that control over the memory and control phases is somehow available to the user; we note some proposals from the literature for achieving this, but are otherwise do not assume a particular means of this control.  Even in the absence of control over the phases (even if we can only set the phases once at the time of device manufacturing)  the device can still be of use for the implementation of an optical waveguide crossing using only beamsplitters and phases, but no actual crossing waveguides, as we discuss in a later Section.

This work is organized as follows.  Section \ref{sec:circuit_construction} constructs the multiplier circuit, starting from the basic component models, using these to build an optical switch/Fredkin gate, and finally using these switches to build the multiplier circuit.  We then extend our construction in several ways: to the computation of matrix-vector and matrix-matrix products and to the use of non-binary selector vectors using an optical feedback construction.  Section \ref{sec:discussion} concludes.  The appendices provide background material for modeling open quantum systems with the Gough-James circuit algebra \cite{GoughJames2008,GoughJames2009}.

\section{Circuit construction} \label{sec:circuit_construction}

We first specify the models for the components of our circuit (beamsplitters, phase shifts, coherent inputs) in Section \ref{sec:component_models}, use these to construct an optical switch/Fredkin gate in Section \ref{sec:MZ}, and then proceed to the optical vector inner product circuit in Section \ref{sec:multiplier}.  Section \ref{sec:matrix-vector_matrix-matrix} extends the construction to the computation of matrix-vector and matrix-matrix products.  Section \ref{sec:construction_feedback} considers an alternate means of computing vector inner products --- modifying the previous construction with an optical feedback loop --- and extends this feedback construction to non-binary selector vectors.


\subsection{Component models} \label{sec:component_models}

We rely upon the formalism of Gough and James \cite{GoughJames2008,GoughJames2009} for modeling circuits of open quantum systems interacting via coherent fields.  This framework is an extension of earlier work on cascaded interacting quantum systems by \cite{HudsonParthasarathy1984,Carmichael993,Gardiner1993,Barchielli2006}; for a brief summary in sufficient detail to repeat our computations or code up a simulation, see Appendix \ref{app:GoughJames_circuit_algebra}.  

For our immediate purposes, we recall that an open quantum system coupled to $n$ incoming and $n$ outgoing optical field modes is parametrized by an ``SLH triplet'' $(\bS,\bL,H)$, where $\bS$ is a $n\times n$ unitary matrix describing the scattering of incoming to outgoing field modes, $\bL$ is a $n\times 1$ vector describing the coupling of the incoming modes to any internal degrees of freedom the system might have, and $H$ is the system Hamiltonian.  The Gough-James circuit algebra provides circuit composition rules to compute SLH triplets for open quantum systems arranged in series, in parallel, and with feedback loops.  The SLH pieces appear in the quantum optical master equation (\ref{eq:master_eq}) for the time evolution of the density matrix for any internal degrees of freedom.

In this section, we shall work only with passive optical components whose models have no internal degrees of freedom --- beamsplitters and constant phase shifts --- that are driven by coherent inputs.  In this setting, the only thing to do is to multiply the scattering matrices of the individual components to obtain an overall scattering matrix for our circuit.    Section \ref{sec:construction_feedback} describes a circuit construction involving feedback, which is perhaps a more intuitive way to construct a multiplier/selector circuit, but involves applying the slightly more complicated feedback operation.

\subsubsection*{Phase shift}

The simplest model in our component toolbox is an optical phase shift --- imparted by a delay line, reflection from a mirror, or some other means.  We shall denote by $\Phi_\phi$ the a component that imparts a phase shift of $\phi$ in an optical path (of course, this is only meaningful if we at some point compare phases of different paths): 
\beq
\Phi_\phi = \left(\bS=e^{i\phi},\bL = 0, H = 0\right)
\eeq
where $\bL$ and $H$ are trivial since there are no internal degrees of freedom to include.  A phase shift of $\phi_1$ followed on the same optical path by a phase shift of $\phi_2$ is denoted by $\Phi_{\phi_2} \lhd \Phi_{\phi_1} = \Phi_{\phi_1+\phi_2}$ and has scattering matrix $\bS = e^{i\left(\phi_1+\phi_2\right)}$.

\subsubsection*{Beamsplitter}

The beamsplitter mixes two incoming fields into two outgoing fields via a rotation matrix\footnote{While the scattering matrix components for a general beamsplitter can be complex, the rotation matrix beamsplitter specified above is sufficient for our needs.} with angle $\theta$.  We denote this component by $B_\theta$:
\beq
B_\theta = \left(\bS = \left(\begin{array}{cc} \cos \theta & -\sin \theta \\ \sin \theta & \cos \theta \end{array}\right),\bL = \left(\begin{array}{c} 0 \\ 0 \end{array}\right), H = 0\right)
\eeq

Like the phase shift component, the beamsplitter has no internal degrees of freedom that we model and so has trivial $\bL$ and $H$.  Putting two beampsplitters in series --- using both outputs of one beamsplitter as inputs for the next --- yields a component equivalent to another beamsplitter: $B_{\theta_2} \lhd B_{\theta_1} = B_{\theta_1 + \theta_2}$.  Setting $\theta = \pm \pi/4$ yields the useful case of a ``50/50'' beamsplitter that splits power incoming to one port equally between the two outputs.



\subsubsection*{Coherent input}

The circuit constructions we present in the following Sections are to be driven by incoming coherent fields.  Because these constructions include only passive components (those that have trivial coupling vector $\bL$ and Hamiltonian $H$, like beamsplitters and phase shifts), they can be described entirely in terms of a circuit scattering matrix $\bS$.  When driven by $n$ coherent inputs with amplitude vector $\vec{\alpha} = (\alpha_1,\ldots,\alpha_n)$, we simply multiply $\bS \vec{\alpha}$ to derive the ``driven'' circuit.  Appendix \ref{app:coherent_input} summarizes the way in which coherent inputs live in the SLH framework and how they appear in the master equation (\ref{eq:master_eq}) for the driven circuit.  The discussion below proceeds entirely in terms of scattering matrices, but we remember to drive the circuits with a nonzero amplitude in order for them to do anything useful.





\subsection{The basic motif}\label{sec:MZ}

\begin{figure}[]
\begin{center}
\includegraphics[width=3.5in]{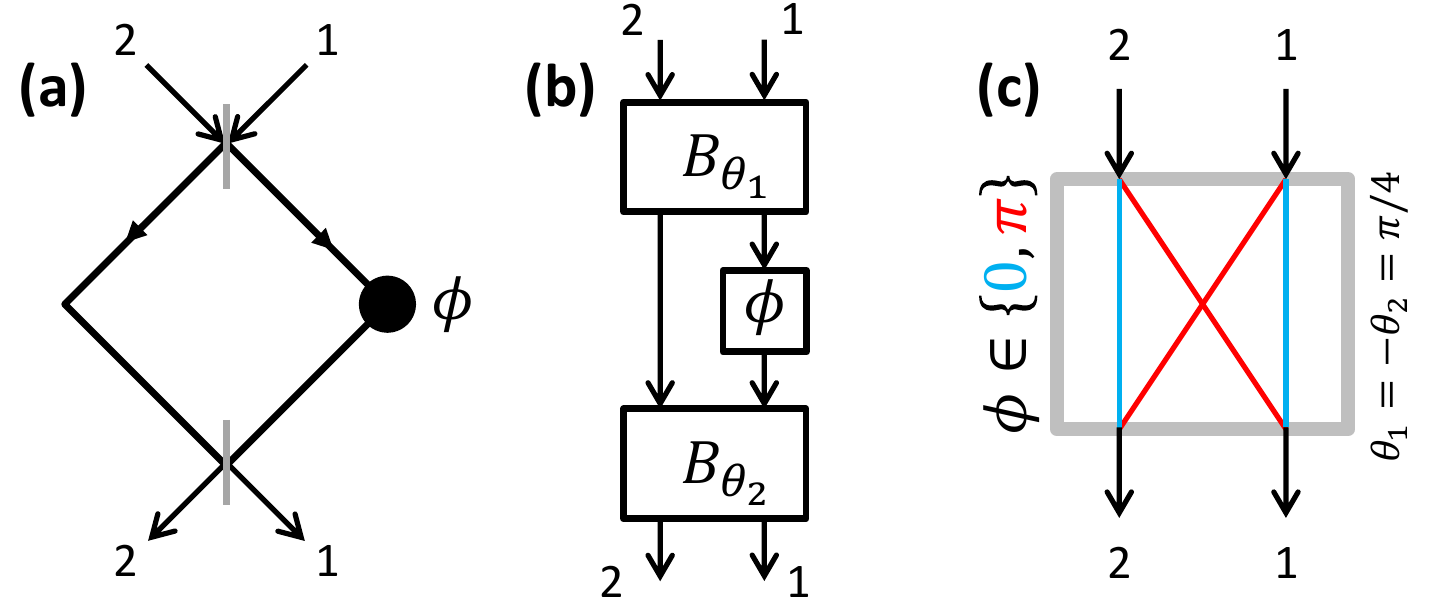}
\caption{The Mach-Zehnder interferometer used as an optical switch / Fredkin gate.  (a) Optical paths, beamsplitters shown in gray.  (b) Music score notation for easy parsing.  (c)  When the beamsplitters are 50/50 ($\theta_1 = -\theta_2 = \pi/4$) and we restrict $\phi\in\{0,\pi\}$, the device is an optical switch / Fredkin gate, switching the inputs conditional on the phase $\phi$.}
\label{fig:Fredkin_gate}
\end{center}
\end{figure}

The basic circuit motif we use to construct the optical multiplier circuit in Section \ref{sec:multiplier} is  the optical equivalent of a switch or Fredkin gate, drawn in Figure \ref{fig:Fredkin_gate}.  This component is a Mach-Zehnder interferometer: Two inputs  mix on beamsplitter  $B_{\theta_1}$.  One leg of the interferometer undergoes a phase shift of $\phi$ relative to the other leg --- let's call this the ``control phase.''  The two legs then mix on a second beamsplitter $B_{\theta_2}$ to produce two outputs.  The SLH model for this device whose two inputs are driven by coherent input fields with magnitudes $(\alpha_1,\alpha_2)$ is given by  
\begin{align}
\text{MZ}_{\theta_1,\theta_2,\phi} & =  B_{\theta_2} \lhd \left(\Phi_{\phi} \boxplus I_1\right) \lhd B_{\theta_1} \\
& = \left(\bS^{\text{MZ}}_{\theta_1,\theta_2,\phi},\bL^{\text{MZ}} = \mathbf{0}, H^\text{MZ} = 0\right) \label{eq:MZ_general_parameters}
\end{align}
where $I_1$ denotes the trivial ``no component'' (with identity scattering matrix) and 
\beq
\bS^{\text{MZ}}_{\theta_1,\theta_2,\phi} = R(\theta_2)\left(\begin{array}{cc}e^{i\phi} & 0 \\ 0 & 1\end{array}\right)R(\theta_1) \label{eq:S_MZ_general_parameters}
\eeq
where $R(\theta)$ denotes the $2\times 2$ rotation by $\theta$ matrix.


To obtain an optical switch, we can restrict the control phase $\phi\in\{0,\pi\}$ and use 50/50 beamsplitters with mixing angles $\theta_1 = -\theta_2 = \pi/4$.  Then the scattering matrix for the above expression (\ref{eq:S_MZ_general_parameters}) evaluates to:
\begin{align}
\bS^{\text{MZ}}_{\pi/4,-\pi/4,\phi} & = \frac{1}{2}\left(\begin{array}{cc} 1+e^{i\phi} & 1-e^{i\phi} \\ 1-e^{i\phi} & 1+e^{i\phi}\end{array}\right) \\
&= \left\{\begin{array}{rl}\left(\begin{array}{cc} 1 & 0 \\ 0 & 1\end{array}\right) :& \phi = 0 \\ \left(\begin{array}{cc} 0 & 1 \\ 1 & 0\end{array}\right) :& \phi = \pi \end{array}\right. \label{eq:S_MZ_phi_0_pi}
\end{align}
so when the control phase $\phi= \pi$ ($ = 0$), the device switches (does not switch) its inputs.

This matches the description of a Fredkin gate, with the control phase $\phi$ playing the role of the control bit that either switches or does not switch the two inputs.  The Fredkin gate is a universal logic gate --- any Boolean function can be implemented using only this element --- so it is a useful component to demonstrate using only the Mach-Zehnder interferometer.  An ``optics inspired'' logic architecture consisting of such devices was proposed in \cite{HardyShamir07}.  Our recent work \cite{PavlichinMabuchi13} implements an error-correcting circuit consisting entirely of these devices.  

\subsubsection{Controlling the control phase}\label{sec:controlling_control_phase}

Of course, to actually put our circuit to use we must be able to control the ``control'' phase $\phi$ somehow, hopefully using yet more optical fields (rather than some other, slower mechanism like changing path lengths mechanically).  One idea, presented in \cite{Mabuchi2009}, is to use an atom-cavity system to stand in for the interferometer.  The atom-cavity dynamics can be engineered in such a way that the phase $\phi \in \{0,\pi\}$ corresponds to the state of the atom, which spends most of the time in one of two nearly degenerate ground states; the atomic state in turn can be ``flipped'' through a coupling to another incoming field, yielding an optical analogue of the set/reset flip-flop latch or switch.  

In the construction of our optical selector/inner product computer in Section \ref{sec:multiplier} we do not assume any particular scheme for setting the control phase $\phi$, but assume that this control is available.  

\subsubsection{Waveguide crossing}\label{sec:waveguide_crossing}

Even in the absence of the ability to dynamically change the control phase $\phi$, the circuit of Figure \ref{fig:Fredkin_gate} is useful in implementing a waveguide crossing.  If we fix the control phase $\phi=\pi$ (perhaps by making one leg of the interferometer longer, or by including an extra reflection in one leg) and choose the beamsplitter angles $\theta_1 = -\theta_2 = \pi/4$, the device always switches (see (\ref{eq:S_MZ_phi_0_pi})) its two inputs to form the outputs --- equivalent to a crossing.  Crossing waveguides reliably (without cross leakage or losses) may be difficult to achieve for integrated nanophotonic systems, so it might be worth the trouble to replace a crossing with two beamsplitters and a phase shift.  The beamsplitters might be constructed by routing two waveguides close enough to each other to be evanescently coupled, or by placing an empty cavity between the two waveguides\footnote{This cavity must be double-sided, resonant with the frequency of the incoming coherent fields, and have equal cavity decay rates from both sides.  The cavity would deviate from beamsplitting behavior by distorting time-varying inputs (by low-pass filtering them), making for a possibly lousy ``beamsplitter,'' but letting us avoid a waveguide crossing!}.  Two proposals for a nanophotonic beamsplitter are \cite{BayindirTemelkuranOzbay2000,Tao_etal2005}.









\subsection{The optical selector/inner product circuit} \label{sec:multiplier}

\begin{figure}[]
\begin{center}
\includegraphics[width=3.5in]{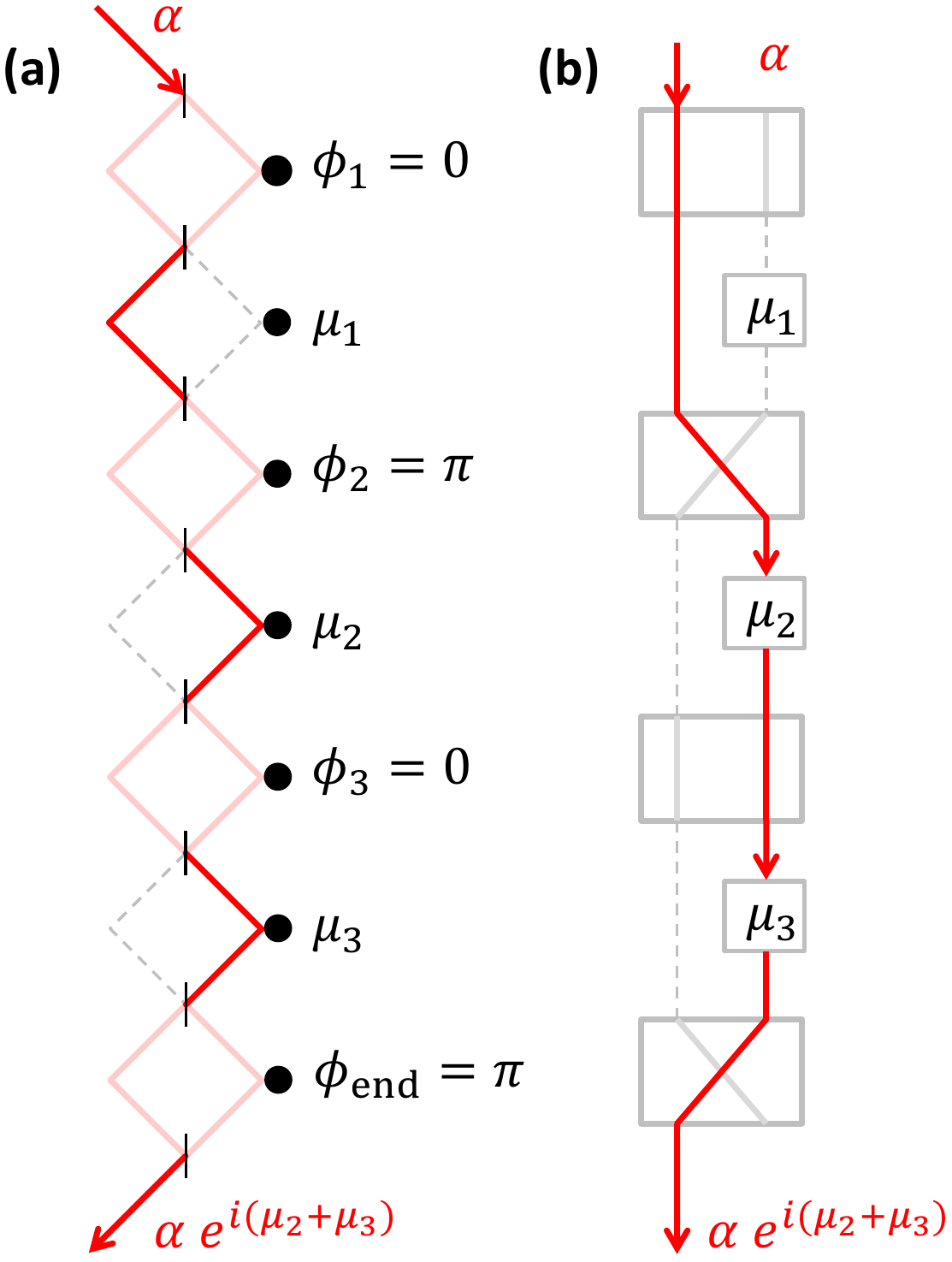}
\caption{(a) Optical multiplier/selector circuit with control phase $\vec{\phi}$ settings indicated on the figure.  Memory phases $\mu_2$ and $\mu_3$ are read and $\mu_1$ is bypassed.  (b) Music-score notation for easy symbolic parsing.}
\label{fig:multiplier}
\end{center}
\end{figure}

We are now ready construct the optical phase selector circuit.  Figure \ref{fig:multiplier} shows this construction: (left panel) a single coherent input with amplitude $\alpha$ (the other input is not used) is incident upon a staircase of $2(n+1)$ 50/50 beamsplitters (the beamsplitter mixing angles alternate $\pi/4,-\pi/4,\pi/4,\ldots$), so that there is a left and a right optical path, crossing at each beamsplitter.  The left and right optical paths between consecutive beamsplitters undergo a relative phase shift named $\phi_1,\mu_1,\phi_2,\mu_2,\ldots,\phi_n,\mu_n,\phi_\text{end}$.  While the ``control'' phases $\phi_i$ and $\phi_\text{end}$ are restricted to the values $0$ or $\pi$, the ``memory'' phases $\mu_i\in[0,2\pi)$ ($\mu$ for $\mu$emory) are arbitrary.  We can imagine viewing each consecutive pair of beamsplitters as a Mach-Zehnder interferometer with control phase shift $\phi$, as discussed in section \ref{sec:MZ} (thus drawn on the right panel).  

The idea is: the control phase shifts $\vec{\phi}=(\phi_i)\in\{0,\pi\}^n$ route the optical input field to either read or not read the $\vec{\mu}=(\mu_i)\in[0,2\pi)^n$ stored memory phase shifts.  In the case of Figure \ref{fig:multiplier}, only $\mu_2$ and $\mu_3$ are read, while $\mu_1$ is bypassed, so the output phase is $\mu_2+\mu_3$.  The final control phase $\phi_\text{end}\in\{0,\pi\}$ is set so that the output exits the device on the left side.  

Our beamsplitter stack conserves optical power and imparts only a phase shift upon the input optical field: the device has only two inputs and two outputs, of which only the left input is used; the final control phase $\phi_\text{end}$ is set so that all of the light leaves from the left output.  Our model does not include optical propagation losses along the beam paths, which could be modeled as extra ``outputs,'' as in \cite{Sarma_etal13}. 

Suppose the input is a coherent field with complex amplitude $\alpha$; then the output of the last beamsplitter is phase-shifted to $\alpha e^{i \mu_\text{out}}$.  What is the relationship between the output phase $\mu_\text{out}$, the control phases $\vec{\phi}$, and the memory phases $\vec{\mu}$?  With some thought we find that the phase of the output field is given by a dot product with a binary ``selector'' vector $\vec{s} \in \{0,1\}^n$:
\begin{align}
\mu_\text{out} & = \vec{\mu} \cdot \vec{s} \\
& = \vec{\mu} \cdot \left(L \vec{\phi}\right) \label{eq:mult_phase} \\
& = \left(\begin{array}{ccc} \mu_1 & \cdots & \mu_n \end{array}\right) \left(\begin{array}{ccc} 1/\pi & 0 & 0 \\ \vdots & \ddots & 0  \\ 1/\pi & \cdots & 1/\pi\end{array}\right) \left(\begin{array}{c} \phi_1 \\ \vdots \\ \phi_n \end{array}\right)
\end{align}
where $L=(L_{ij})$ is the $n\times n$ lower-triangular matrix of $1/\pi$s:
\beq
L_{ij} = \frac{1}{\pi}\ \mathbb{I}_{i \leq j} \label{eq:lower_triangular_mat}
\eeq
To ensure that the output is always on the left side of the device we must set the final control phase $\phi_\text{end}$: 
\beq
\phi_\text{end} = \sum_{i=1}^n \phi_i 
\label{eq:mult_control_phase_end}
\eeq
The arithmetic in (\ref{eq:mult_phase}) and (\ref{eq:mult_control_phase_end}) is modulo $\pi$.

Suppose we start with a binary selector vector $\vec{s}\in\{0,1\}^n$ specifying which of the memory phases we want to read out.  How should we set the control phases $\vec{\phi}\in\{0,\pi\}^n$ to achieve this selection?  Inverting the expression in (\ref{eq:mult_phase}) we find
\beq
\vec{\phi} =L^{-1} \vec{s} = \mathit{\Gamma} \vec{s} \label{eq:phi_from_selector}
\eeq
where $L^{-1} = \mathit{\Gamma} =(\mathit{\Gamma}_{ij})$ is the $n \times n$ double-band-diagonal matrix
\beq
\mathit{\Gamma}_{ij} = \pi\left(\mathbb{I}_{i = j} - \mathbb{I}_{i = j + 1}\right) \label{eq:double_band_diagonal_mat}
\eeq
Finally, we must again choose $\phi_\text{end}$ to satisfy (\ref{eq:mult_control_phase_end}) to ensure that all of the light comes out the left port\footnote{Since $\phi_\text{end}$ is a linear combination of the $\phi_i$s, we could include its computation in (\ref{eq:phi_from_selector}) with an augmented $(n+1)\times (n+1)$ matrix $\mathit{\Gamma}$ and a dummy $(n+1)$-st memory phase that never gets read, but choose to keep $\phi_\text{end}$ separate to avoid complicating our notation.}.

We now have an optical circuit that can read out arbitrary selections of stored memory phases through an appropriate setting of the control phases.  We next extend the construction slightly to make an optical matrix-vector and matrix-matrix multiplier.

\subsection{Optical matrix-vector and matrix-matrix multiplier} \label{sec:matrix-vector_matrix-matrix}

\begin{figure}[]
\begin{center}
\includegraphics[width=3.4in]{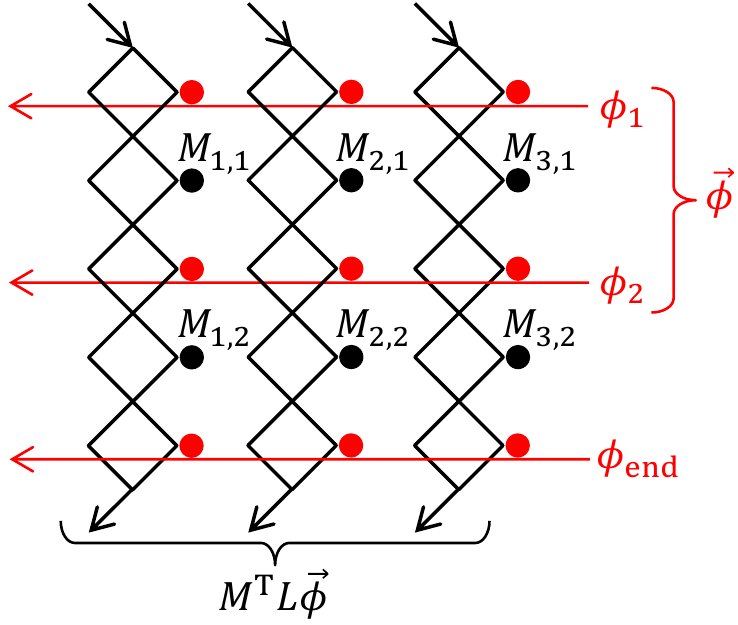}
\caption{Optical arbitrary-matrix-binary-vector multiplication circuit obtained by repeating the construction of Figure \ref{fig:multiplier} (a beamsplitter is implicit at every waveguide crossing).  The control phases $\vec{\phi}\in\{0,\pi\}^n$ and $\phi_\text{end}$ are distributed to three copies of the multiplier circuit.  The $n \times m$ matrix $M\in[0,2\pi)^{nm}$ of memory phases determine the output phases via the relation (\ref{eq:phases_selector_relationship_matrix_mult}).}
\label{fig:matrix_multiplication}
\end{center}
\end{figure}

Now that we can compute the dot product of a vector of stored memory phases $\vec{\mu}$ with a binary selector vector, we can straightforwardly extend the construction to compute matrix-vector products by simply placing several copies of this circuit next to each other, as shown in Figure \ref{fig:matrix_multiplication}.  We can imagine extending the circuit to a matrix-vector product for a matrix of either binary control phases or arbitrary memory phases (or both, for a matrix-matrix product), but restrict attention below initially to the matrix of arbitrary memories case.  



To realize this circuit we must somehow provide a copy of the control phases vector $\vec{\mu}$ to each vector-vector inner product subcircuit (each column of beamsplitters in Figure \ref{fig:matrix_multiplication}).  One way to achieve this would be to use the photonic latch of \cite{Mabuchi2009}, discussed in more detail in Section \ref{sec:controlling_control_phase}.  Then we could encode the control phase information in the state of a set of incoming fields (drawn in red in Figure \ref{fig:matrix_multiplication}) that would in turn drive the control phases $\phi$ into the desired $0$ or $\pi$ state.  As before, we assume that some way to set the control phases is available and proceed without assuming a particular physical implementation.  If the control phases are distributed via coherent fields in waveguides, then we can see in Figure \ref{fig:matrix_multiplication} that there are crossings with the waveguides implementing the optical multiplication; a possible way of handling these crossings using yet more interferometers is outlined in Section \ref{sec:waveguide_crossing}.


Let's characterize this circuit.  Suppose we have $m$ memory phase vectors of length $n$ that we arrange into the $n \times m$ matrix $M$:
\beq
M = \left(\begin{array}{ccc}\vec{\mu}_1 & \cdots & \vec{\mu}_m\end{array}\right) \in [0,2\pi)^{n m}
\eeq
Then the $m \times 1$ vector $\vec{\mu}_\text{out}$ of output field phases is given by the matrix version of (\ref{eq:mult_phase}):
\beq
\vec{\mu}_\text{out}\T = M\T \vec{s} = M\T L \vec{\phi}
\eeq
where $\vec{s}$ is the $n \times 1$ binary selector vector related to the control phase vector $\vec{\phi}$ by (\ref{eq:phi_from_selector}).

Suppose we extend this construction to a matrix-matrix multiplier (by making parallel copies of the matrix-vector multiplier of Figure \ref{fig:matrix_multiplication}) for $k$ control phase vectors of length $n$ that we arrange into the $n \times k$ matrix $\mathit{\Phi}$:
\beq
\mathit{\Phi} = \left(\begin{array}{ccc} \vec{\phi}_1 & \cdots & \vec{\phi}_k\end{array}\right)\in\{0,\pi\}^{nk}
\eeq
Then the $m \times k$ matrix of output phases $M_\text{out}$ is the matrix product of $M$ and a binary $n \times k $ selector matrix $S\in\{0,1\}^{nk}$
\beq
M_\text{out} = M\T S = M\T L \mathit{\Phi}
\eeq
where $\mathit{\Phi}$ and $S$ are related by the matrix version of (\ref{eq:phi_from_selector}):
\beq
\mathit{\Phi} = \pi \mathit{\Gamma} S \label{eq:phases_selector_relationship_matrix_mult}
\eeq
where $\mathit{\Gamma}$ is the double-band-diagonal matrix given in (\ref{eq:double_band_diagonal_mat}).  Analogously to (\ref{eq:mult_control_phase_end}), the $k \times 1$ vector of tail phases $\vec{\phi}_\text{end}\in\{0,\pi\}^k$ is given by
\beq
\vec{\phi}_\text{end} = \mathit{\Phi}\T \mathbf{1}_{n \times 1}
\eeq
where $\mathbf{1}_{n \times 1}$ is the vector of all $1$s.

\subsection{Constructions with feedback} \label{sec:construction_feedback}

In this section we point out an alternate way to construct the optical dot product circuit.  This alternate construction involves feedback, which might be difficult to engineer, but simplifies the circuit in some ways.  

\begin{figure}[]
\begin{center}
\includegraphics[width=3in]{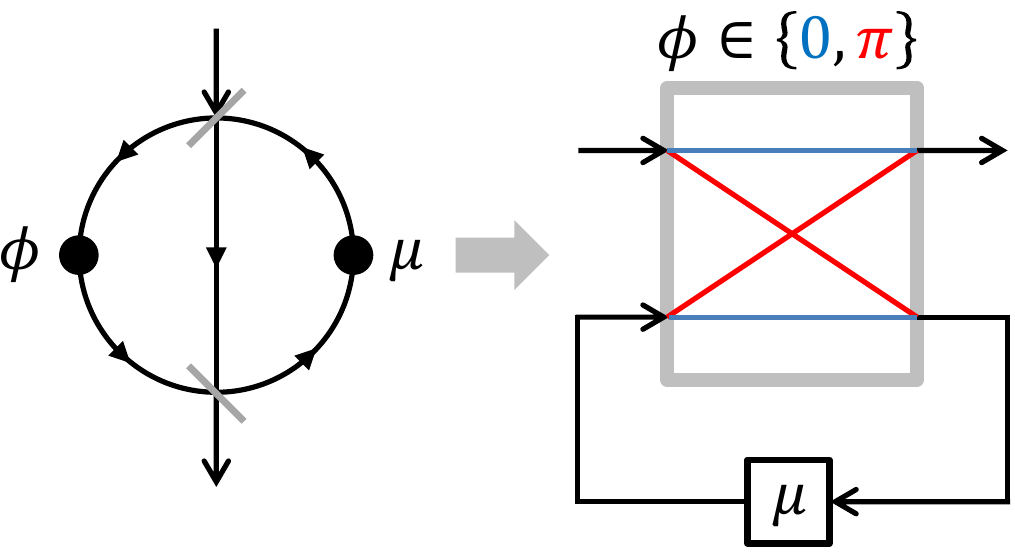}
\caption{(left) Portion of optical phase selector with feedback with a single memory phase $\mu$ and control phase $\phi$.  (right) When the control phase $\phi$ is restricted to $\phi \in \{0,\pi\}$, the input field loops back through memory phase $\mu$ once (for $\phi = \pi$) or zero times (for $\phi = 0$) before exiting the device.}
\label{fig:multiplier_fb}
\end{center}
\end{figure}

Figure \ref{fig:multiplier_fb} shows a sub-unit of the optical multiplier/selector corresponding to a single control phase $\phi\in\{0,\pi\}$ and a single memory phase $\mu$.  The circuit consists of a Mach-Zehnder interferometer with control phase $\phi$ of one optical path relative to the other (just as in the construction of Section \ref{sec:multiplier}), but now one of the outputs of the second beamsplitter is fed back to one of the inputs of the first beamsplitter after picking up an extra memory phase $\mu$.  We use $B_{\pi/4}$ and $B_{-\pi/4}$ 50/50 beamsplitters, in that order.

Denote the SLH circuit model for this device by $F_{\phi,\mu}$:
\begin{align}
F_{\phi,\mu} & = \left[(\Phi_\mu \boxplus I_1) \lhd B_{-\pi/4} \lhd (\Phi_\phi \boxplus I_1) \lhd B_{\pi/4}\right]_{1 \rightarrow 1} \\
& = \left(\bS_{\phi,\mu},\bL = \mathbf{0}, H = 0\right)
\end{align}
where $[\cdot]_{1 \rightarrow 1}$ denotes the feedback of output port 1 back to input port 1.  Applying the feedback transformation specified in Appendix \ref{app:GoughJames_circuit_algebra} (\ref{eq:feedback_operation}), we find that the $1 \times 1$ scattering matrix is
\beq
\bS_{\phi,\mu} = \frac{1+e^{i\phi}-2e^{i(\phi+\mu)}}{2-e^{i\mu}+e^{i(\phi+\mu)}} \label{eq:S_feedback_mult}
\eeq

\subsubsection{Restricting the control phase to $0$ or $\pi$: Switch}

We evaluate (\ref{eq:S_feedback_mult}) for control phase $\phi = 0$ (resp. $\phi = \pi$) to find $\bS_{0,\mu} = 1$ (resp. $\bS_{\pi,\mu} = e^{i\mu}$), so as for the selector circuit of Section \ref{sec:multiplier}, the control phase determines whether the memory phase is read or not.  As shown in the right pane of Figure \ref{fig:multiplier_fb}, when $\phi = 0$, the incoming field bypasses the memory phase and exits the device; when $\phi = \pi$, the incoming field loops back through the memory phase $\phi$ before exiting the circuit.  

Using feedback simplifies the circuit in several ways.  First, if we string $n$ such devices together, there is a simpler relationship between the control phase vector $\vec{\phi}$ and the binary selector vector $\vec{s}$ for the memory phases.  The output phase is simply
\beq
\mu_\text{out} = \arg \bS_{\phi,\mu} = \vec{\mu}\cdot\vec{s} =\frac{1}{\pi}\vec{\mu}\cdot\vec{\phi}
\eeq
when $\phi\in\{0,\pi\}$.  Thus $\vec{s} = \frac{1}{\pi}\vec{\phi}$ and there is no longer a triangular or double-band-diagonal matrix to keep track of when converting between control phase vectors and binary selector vectors as in Section \ref{sec:multiplier}.  Second, the device of Figure \ref{fig:multiplier_fb} has only one output port, so there is no longer a need to keep a separate tail control phase $\phi_\text{end}\in\{0,\pi\}$ (\ref{eq:mult_control_phase_end}) to ensure all of the light exits from a particular output port.

These nice features come at the expense of having a feedback loop.  This controlled feedback motif was used in our earlier work \cite{PavlichinMabuchi13} (where instead of a phase $\mu$, an attenuating beam dump was placed in the feedback loop).  



\subsubsection{Arbitrary control phase}

Letting the control phase $\phi$ be arbitrary ($\phi\in[0,2\pi)$) rather than binary enables us to partially select an encoded memory phase.  This could be useful for computational tasks requiring arbitrary vector-vector inner products, such as statistical classification or perceptron training. 

For a single feedback-style selector of Figure \ref{fig:multiplier_fb}, the phase shift imparted upon the outgoing coherent field is given by $\mu_\text{out} = \arg \bS_{\phi,\mu}$, where $\bS_{\phi,\mu}$ is given in (\ref{eq:S_feedback_mult}).  It isn't obvious what such a relationship could be useful for, other than resulting in some nonlinear coupling between the output phase and memory phase conditional on the control phase.  

\begin{figure}[]
\begin{center}
\includegraphics[width=2in]{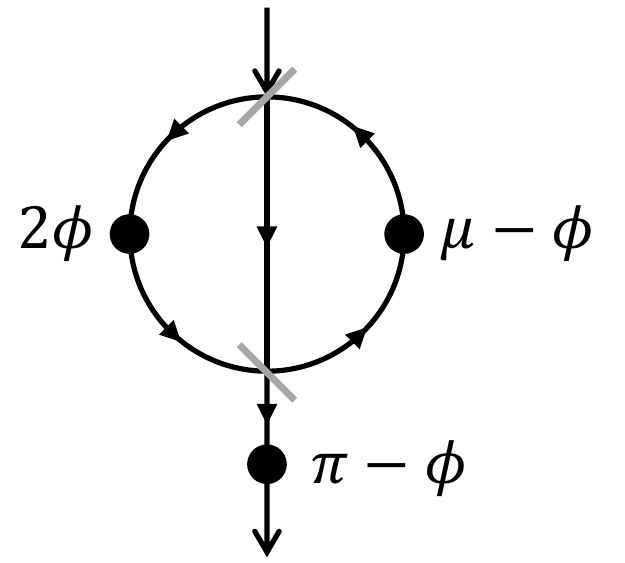}
\caption{Modified feedback circuit of Figure \ref{fig:multiplier_fb} that computes the product $\mu_\text{out} \approx \mu \cot^2(\phi/2)$ for arbitrary (non-binary) control phase $\phi$ and small absolute value of memory phase $\mu$.}
\label{fig:fb_nonbinary_control_phase}
\end{center}
\end{figure}

We can obtain a more interpretable system by modifying the feedback circuit somewhat to include two more appearances of the control phase $\phi$, as shown in Figure \ref{fig:fb_nonbinary_control_phase}.  The SLH model is:
\beq
\Phi_{\pi-\phi} \lhd \left[(\Phi_{\mu-\phi} \boxplus I_1) \lhd B_{-\pi/4} \lhd (\Phi_{2\phi} \boxplus I_1) \lhd B_{\pi/4}\right]_{1 \rightarrow 1}
\eeq

\begin{figure}[]
\begin{center}
\includegraphics[width=3.5in]{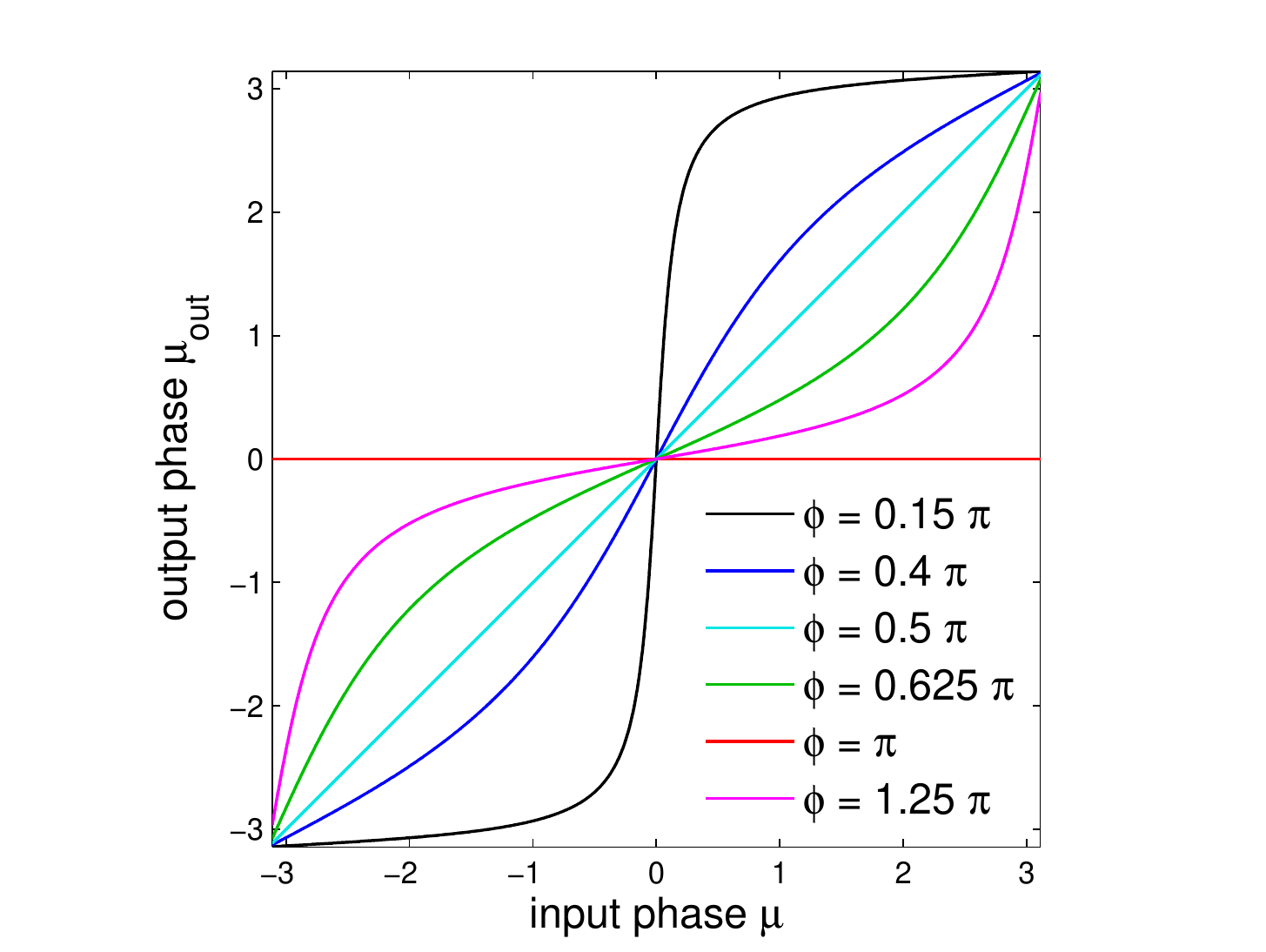}
\caption{Output phase $\mu_\text{out}$ vs memory phase $\mu$ for the feedback circuit of Figure \ref{fig:fb_nonbinary_control_phase} plotted for several values of the control phase $\phi$.  For small $|\mu|$, we have $\mu_\text{out} \approx \mu \cot^2(\phi/2)$.}
\label{fig:FB_phase_fig}
\end{center}
\end{figure}

We find the output phase $\mu_\text{out}$ for this circuit to be:
\begin{align}
\mu_\text{out} & = \arctan \left(\frac{4 \sin^2 \phi \sin \mu}{2(3+\cos(2\phi))\cos \mu - 8 \cos \phi}\right) \\
& = \mu \cot^2(\phi/2) + O(\mu^3) \label{eq:mu_out_fb_series_mu}
\end{align}
Thus for small values of $\mu$ (for $|\mu \cot^2(\phi/2)| \ll \pi$), the circuit of Figure \ref{fig:fb_nonbinary_control_phase} computes the product $\mu_\text{out} \approx \mu \cot^2(\phi/2)$, where now $\phi$ is arbitrary (rather than binary).  Figure \ref{fig:FB_phase_fig} plots the memory-phase-to-output-phase ``transfer function'' for several values of the control phase $\phi$.  We see that now the case $\phi=\pi$ (resp. $\phi = \pi/2$) corresponds to $\mu_\text{out} = 0$ (resp. $\mu_\text{out} = \mu$).  

For values of $\phi$ near $\pi/2$, we can expand (\ref{eq:mu_out_fb_series_mu}) in powers of $(\phi-\pi/2)$ to obtain:
\beq
\mu_\text{out}  = \mu\left(1-(\phi-\pi/2) + O((\phi-\pi/2)^2)\right) + O(\mu^3)
\eeq


\section{Discussion}\label{sec:discussion}

We have presented an optical circuit that computes the inner product of a binary selector vector with an arbitrary vector of phases, and encodes the output in the phase of an outgoing coherent field.  This operation could be useful for the construction of an optical random access memory, or for any computational routine that involves modular vector arithmetic.  The additional constructions we describe for matrix-vector and matrix-matrix products, and for the use of non-binary selector vectors via an optical feedback construction, extend the applicability of this circuit to algorithms that rely upon these more general operations.  

The circuit makes natural use of a collection of optical devices: signals are encoded in optical phases imparted upon a coherent field and are routed with interferometers.  While the proposed device itself is not surprising in that it mimics an electronic circuit made of Fredkin gates, our construction is a demonstration of the computational use of optical circuits and of the design methodology enabled by recently-developed mathematical tools for describing such circuits.  


\appendix

\section{Circuit algebra} \label{app:GoughJames_circuit_algebra}

The Gough-James circuit algebra \cite{GoughJames2008,GoughJames2009} parametrizes an open quantum system coupled to $n$ external field modes with a triplet $(\bS,\bL,H)$, where $H$ is the Hamiltonian for the system's internal degrees of freedom, $\bL$ is a $n \times 1$ coupling vector corresponding to the interactions of external field modes with internal degrees of freedom, and $\bS$ is a $n \times n$ unitary matrix describing the scattering of incoming to outgoing field modes.  The entries of $(\bS,\bL,H)$ are in general operator-valued [example below].

The density matrix $\rho$ for the system's internal degrees of freedom involves in time according to the master equation:
\beq
\dot{\rho}_t = -i  [H,\rho_t] + \sum_{i=1}^n \left(L_i \rho_t L_i^\dagger - \frac{1}{2}\{L_i^\dagger L_i ,\rho_t\}\right) \label{eq:master_eq}
\eeq
where $[A,B] = AB-BA$, $\{A,B\} = AB+BA$, and $\dagger$ denotes conjugation.  The scattering matrix $\bS$ does not appear in (\ref{eq:master_eq}); when we connect two open quantum systems in series below, the scattering matrix of one component does appear in the effective $\bL$ for the whole circuit, and in this way can enter the master equation.

The Gough-James circuit algebra provides composition rules to derive $(\bS,\bL,H)$ triplets for open quantum systems arranged in a circuit.  Two open quantum systems $G_1 = (\bS_1,\bL_1,H_1)$ and $G_2 = (\bS_2, \bL_2, H_2)$ can be arranged in series so long as both are coupled to the same number of external field modes.  Connecting the outputs of $G_1$ to the inputs of $G_2$ corresponds to the series product
\beq
G_2 \lhd G_1 = \left(\mathbf{S}_2 \mathbf{S}_1,\ \mathbf{S}_2 \mathbf{L}_1 + \mathbf{L}_2,\ H_1 + H_2 + \Im\left(\mathbf{L}_2^\dag \mathbf{S}_2 \mathbf{L}_1\right)\right) \label{eq:series_product}
\eeq
where $\Im(X) = \frac{1}{2i}\left(X-X^\dag\right)$ denotes the imaginary part.

The concatenation product corresponds to treating two non-interacting systems $G_1$ and $G_2$, coupled to $n_1$ and $n_2$ external field modes, respectively, and viewing them as a single open quantum system coupled to $n_1 + n_2$ external field modes:
\beq
G_1 \boxplus G_2 = \left(\left(\begin{array}{cc} \mathbf{S}_2 & \mathbf{0} \\ \mathbf{0} & \mathbf{S}_1 \end{array}\right), \left(\begin{array}{c}\mathbf{L}_1 \\ \mathbf{L}_2 \end{array}\right), H_1 + H_2\right) \label{eq:concatenation_product}
\eeq

Finally, the feedback operation denoted by $\left[G\right]_{k \rightarrow l}$ for a system $G$ coupled to $n$ external field modes corresponds to feeding back the $k$-th output of $G$ to the $l$-th input, yielding a system coupled to $n-1$ external field modes:
\beq
[G]_{k\rightarrow l} = (\bS_\text{fb}, \bL_\text{fb}, H_\text{fb}) \label{eq:feedback_operation}
\eeq
where
\begin{align}
\bS_\text{fb} & = \bS^{\cancel{[k,l]}} + \bS^{\cancel{[k]}}_{:,l} \left(1-S_{k,l}\right)^{-1} \bS^{\cancel{[l]}}_{k,:} \\
\bL_\text{fb} & = \bL^{\cancel{[k]}} + \bS^{\cancel{[k]}}_{:,l} \left(1-S_{k,l}\right)^{-1} L_k \\
H_\text{fb} & = H + \Im\left[\left(\sum_{j=1}^n L_j^\dag S_{j,l}\right)\left(1-S_{k,l}\right)^{-1}L_k\right]
\end{align}
where $\cancel{[k,l]}$ (resp. $\cancel[k]$) denotes the matrix (resp. vector) obtained by removing row $k$ and column $l$ (resp. entry $k$) and $\bS_{:,l}$ (resp. $\bS_{k,:}$) denotes taking the $l$-th column (resp. $k$-th row) of matrix $\bS$.




\section{Coherent input} \label{app:coherent_input}





We represent driving optical components with a coherent input field with another ``component.''  We denote by $W_{\vec{\alpha}}$ a coherent input field with $n\times 1$ complex amplitude $\alpha$, corresponding to the Weyl operator that displaces the $n\times 1$ vacuum input $\vec{0}$ into the coherent state $|\vec{\alpha}\rangle = |\alpha_1,\ldots,\alpha_n\rangle$.  In SLH terms this is
\beq
W_{\vec{\alpha}} = \left(\mathbf{S} = \mathbf{1}_{n\times n},\ \ \ \mathbf{L} = \left(\begin{array}{c}\alpha_1\\ \vdots \\ \alpha_n \\\end{array}\right),\ \ \ H = 0\right)
\eeq
where $\mathbf{1}_{n\times n}$ is the identity matrix.  We drive a component $G$ with coherent input $W_{\vec{\alpha}}$ by arranging them in series: $G \lhd W_{\vec{\alpha}}$.  For our needs, every component (beamsplitter, phase shift) includes only a non-trivial scattering matrix (and trivial coupling vector and Hamiltonian), so computing this series product is a matter of multiplying the scattering matrix of component $G$ by the vector $(\alpha_1,\ldots,\alpha_n)^\text{T}$ to obtain the overall coupling vector $\bL$.  For example, driving a 50/50 beamsplitter with coherent input amplitudes $(\alpha_1,\alpha_2)$ is written:
\beq
B_{\pi/4} \lhd W_{(\alpha_1,\alpha_2)} = \left(\frac{1}{\sqrt{2}}\left(\begin{array}{cc} 1 & -1  \\ 1 & 1\end{array}\right), \frac{1}{\sqrt{2}}\left(\begin{array}{c} \alpha_1 - \alpha_2 \\ \alpha_1 + \alpha_2 \end{array}\right), 0\right)
\eeq
where we applied the series product rule (\ref{eq:series_product}).  This ``driven'' beamsplitter now has a nontrivial coupling vector $\bL$, which appears in the master equation (\ref{eq:master_eq}) for this circuit.

\acknowledgements

This work was supported by DARPA-MTO under award \#N66001-11-1-4106.

\bibliography{refs}   

\begin{thebibliography}{10}

\bibitem{Beau11}
Beausoleil, R.~G., ``Large-scale integrated photonics for high-performance
  interconnects,'' {\em J. Emerg. Technol. Comput. Syst.}~{\bf 7},  6:1--6:54
  (July 2011).

\bibitem{Mill09}
Miller, D., ``Device requirements for optical interconnects to silicon chips,''
  {\em Proceedings of the IEEE}~{\bf 97}(7),  1166--1185 (2009).

\bibitem{PavlichinMabuchi13}
Pavlichin, D.~S. and Mabuchi, H., ``Photonic circuits for iterative decoding of
  a class of low-density parity-check codes,'' {\em pre-print}  (2013).

\bibitem{GoughJames2008}
Gough, J., J. M.~R., ``Quantum feedback networks: Hamiltonian formulation,''
  {\em Commun. Math. Phys.}~{\bf 287},  1109--1132 (2008).

\bibitem{GoughJames2009}
Gough, J., J. M.~R., ``The series product and its application to quantum
  feedforward and feedback networks,'' {\em IEEE Transactions on Automatic
  Control}~{\bf 54},  2530--2544 (2009).

\bibitem{HudsonParthasarathy1984}
Hudson, R.~L. and Parthasarathy, K.~R., ``Quantum ito's formula and stochastic
  evolutions,'' {\em Communications in Mathematical Physics}~{\bf 93},
  301--323 (1984).

\bibitem{Carmichael993}
Carmichael, H.~J., ``Quantum trajectory theory for cascaded open systems,''
  {\em Physical Review Letters}~{\bf 70},  2273--2276 (1993).

\bibitem{Gardiner1993}
Gardiner, C.~W., ``Driving a quantum system with the output field from another
  driven quantum system,'' {\em Physical Review Letters}~{\bf 70},
  2269–--2272 (1993).

\bibitem{Barchielli2006}
Barchielli, A.,  [{\em Open Quantum Systems III: Recent
  Developments}{\nolinebreak\hspace{0.1em}]}, Springer-Verlag, New York, NY
  (2006).

\bibitem{HardyShamir07}
Hardy, J. and Shamir, J., ``Optics inspired logic architecture,'' {\em Optics
  Express}~{\bf 15},  150--165 (2007).

\bibitem{Mabuchi2009}
Mabuchi, H., ``Cavity-qed models of switches for attojoule-scale nanophotonic
  logic,'' {\em Phys. Rev. A}~{\bf 80},  045802 (2009).

\bibitem{BayindirTemelkuranOzbay2000}
Bayindir, M., Temelkuran, B., and Ozbay, E., ``Photonic-crystal-based beam
  splitters,'' {\em Applied Physics Letters}~{\bf 77},  3902 (2000).

\bibitem{Tao_etal2005}
Liu, T., Zakharian, A., Fallahi, M., Moloney, J.~V., and Mansuripur, M.,
  ``Design of a compact photonic-crystal-based polarizing beam splitter,'' {\em
  IEEE Photonics Technology Letters}~{\bf 17},  1435--1437 (2005).

\bibitem{Sarma_etal13}
Sarma, G., Hamerly, R., Tezak, N., Pavlichin, D.~S., and Mabuchi, H.,
  ``Transformation of quantum photonic circuit models by term rewriting,'' {\em
  IEEE Photonics Journal}~{\bf 5},  7500111 (2013).

\end{thebibliography}
\bibliographystyle{spiebib}   

\end{document}